%% file: iccad2011.tex
\documentclass[10pt,conference]{IEEEtran}
   \usepackage{epsfig}
   \usepackage{color}
   \usepackage{latexsym}
   \usepackage{pslatex}
  \usepackage{algorithm}
   \usepackage{algorithmic}
   \usepackage{multirow}

\newtheorem{definition}{Definition}
\newtheorem{lemma}{Lemma}
\newtheorem{theorem}{Theorem}

\begin{document}

\title{Synthesis of Parallel Binary Machines} 
\author{}
\author{Elena~Dubrova, \\
Royal Institute of Technology, IMIT/KTH, 164 46 Kista, Sweden}

\maketitle

\begin{abstract}
Binary machines are a generalization of Feedback Shift Registers (FSRs) in which both, feedback and feedforward, connections are allowed and no chain connection between the register stages is required. In this paper, we present an algorithm for synthesis of binary machines with the minimum number of stages for a given degree of parallelization. Our experimental results show that for sequences with high linear complexity such as complementary, Legendre, or truly random, parallel binary machines are an order of magnitude smaller than parallel FSRs generating the same sequence. The presented approach can potentially be of advantage for any application which requires sequences with high spectrum efficiency or high security, such as data transmission, wireless communications, and cryptography.
\end{abstract}

\begin{keywords}
Feedback shift register, sequences, nonlinear complexity
\end{keywords}

\section{Introduction}

In information theory, it is known that any binary sequence with a finite period
can be generated by a {\em binary machine} shown in Figure~\ref{bin_machine} \cite{Golomb_book}.
An $n$-stage  binary machine consists of an $n$-stage binary register, $n$ updating Boolean functions, and a clock. At each clock cycle, the current values of all stages of the register are synchronously updated to the next values computed by the updating functions. Binary machines can be viewed as a more general version of Feedback Shift Registers (FSRs).

Suppose we would like to construct a binary machine which generates the following 
binary sequence:
\[
A_2 = (0,0,1,1,0,1,1,1,0,0,1,0,1,1,1,0,1,1,0,0).
\]

Since the output of a binary machine equals to the least significant bit
of its current state, any assignment of states $S_2 = (s_0, s_1, \ldots, s_{19})$ 
such that $s_i$ mod $2 = a_i$ results in a binary machine with generates $A$. For example, we can use
\[
S_2 = (0,2,1,3,4,5,7,9,6,8,11,10,13,15,17,12,19,21,14,16)
\]
where even and odd integers are assigned in an increasing order. From $S_2$ we can easily
see how many stages a binary machine should have to generate $A_2$. The largest element of $S_2$ is 21.
We need 5 bits to expand it in binary. Thus, a binary machine generating $A_2$ should have at 
least 5 stages. 

As in the case of traditional Finite State Machines (FSM) synthesis~\cite{MiBS85}, for different state assignments
we usually get different next state functions. The circuit complexity of these functions may vary 
substantially for different state assignments. We can also use the one-hot encoding 
instead of the binary one. 
Then, the number of stages will increase, but
the complexity of functions might decrease in some cases.

\begin{figure}[t!]
\begin{center}
\resizebox{0.99\columnwidth}{!} {\input{binary_machine.xfig.pstex_t}}
\caption{ A binary $n$-stage machine with the degree of parallelization one.}\label{bin_machine}
\end{center}
\end{figure}

Next we describe an intuitive idea behind the algorithm for synthesis of 
parallel binary machines presented in this paper.
Suppose that we use the encoding (00) = 0, (01) = 1, (10) = 2, (11) = 3
to encode the binary sequence $A_2$ from the example above into
the following quaternary sequence:
\[
A_4 = (0,3,1,3,0,2,3,2,3,0).
\]
We can construct a quaternary machine generating $A_4$ (in which the stages of the register can store 4 different values and the updating functions are 4-valued) by choosing a sequence 
of states $S _4 = (s_0, s_1, \ldots, s_{9})$ such that $s_i$ mod $4 = a_i$.
For example, we can assign the states as follows:
\[
S_4 = (0,3,1,7,4,2,11,6,15,8).
\]
Note that the largest element of $S_4$ is 15. We need 2 quaternary digits to represent it.
Thus, we can generate $A_4$ using a quaternary machine with 2 stages (see Figure~\ref{bin_mv}(a)). 
Such a quaternary machine can, in turn, be converted into a binary machine 
by encoding each 4-valued function by a pair of Boolean functions and
by replacing each quaternary stage by two binary stages (see Figure~\ref{bin_mv}(b)).
The resulting 4-stage binary machine generates the same binary sequence $A_{2}$ as in the example above, but
two bits per clock cycle. Note, that is the example above we needed 5 stages to generate $A_2$ one bit per clock cycle. 
So, we constructed a parallel binary machine which has fewer stages than the theoretical lower bound on the 
number of stages in a binary machines generating the same sequence sequentially bit by bit.  

Later in the paper, we show that the number of stages can be reduced even further by using the 8-ary encoding. 
What is even more important, we reduce not only the number of stages, but also the circuit complexity of the updating 
functions.  Our experimental results show that for sequences with high linear complexity such as complementary, Legendre, or truly random, parallel binary machines are an order of magnitude smaller than parallel FSRs generating the same sequence.
Therefore, 
the presented approach can potentially be useful
for any application which requires sequences with high spectrum efficiency or high security.
Such applications include
data transmission, wireless communications, cryptography, and many others~\cite{ZeYWR91,DaJ98,popovich, KrP04}. A particularly attractive application is
encryption and authentication systems for  smartcards 
and Radio Frequency IDentification (RFID) tags.
A low-cost RFID tag can spare only a few hundred gates for security functionality~\cite{rfid_review}.
None of the available cryptographic systems satisfies this requirement at present~\cite{GoB08}.

The rest of the paper is organised as follows. Section~\ref{prel} describes basic notation and definitions used in the sequel.
In Section~\ref{sa1}, we present an algorithm for constructing an $m$-ary
machine with the minimum number of stages generating a given
$m$-ary sequence. In Section~\ref{bin}, we show how $m$-ary
machines can be encoded to generate binary sequences in parallel and demonstrate that such an encoding can be of advantage.
Section~\ref{exp} presents the experimental results.
Section~\ref{con} concludes the paper.

\section{Preliminaries} \label{prel}

Let $M = \{0,1,\ldots,m-1\}$. An $m$-{\em ary sequence} is vector $A_m = (a_0, a_1, \ldots,)$
where $a_i \in M$ for all $i \geq 0$.  

If there exist $k > 0$ and $k_0 \geq 0$
such that $a_i = a_{i+k}$ for all $i \geq k_0$, then $A$ 
is called {\em eventually} (or {\em ultimately}) {\em periodic}. 
If $k_0 = 0$, then $A$ is called {\em purely periodic}, or simply {\em periodic}. 
The least integers $k_0$ and $k$ with this 
property are called {\em pre-period} and {\em period} of the sequence, respectively~\cite{LiKK07}. 

For a multiple-valued function $f: M^n \rightarrow M$, 
the $i$-{\em set of} $f$ is defined by~\cite{dubrova2001x} 
\[
i{\mbox{\em-set}}(f) = \{x \in M^n: f(x) = i\}.
\]
In the binary case, 0-set and 1-set correspond to off-set and on-set of $f$, respectively~\cite{espr}.




An $m$-{\em ary} $n$-{\em stage machine} consists of $n$
$m$-ary storage elements, called {\em stages}. 
Each stage $i \in \{0,1,\ldots,n-1\}$ has an associated {\em state variable} $x_i \in M$ 
which represents the current value of the stage $i$ and an {\em updating function} 
$f_i: M^n \rightarrow M$ which determines how the value of $x_i$ is updated. 

A {\em state} of an $n$-stage machine is a vector of values of its
state variables. At every clock cycle, 
the next state of a machine is determined from its the current state 
by updating the values of all stages simultaneously
to the values of the corresponding $f_i$'s.
 
The {\em degree of parallelization} of an $n$-stage machine is the number 
of stages $p$, $1 < p \leq n$, which are used to produce its output at each clock cycle. 

\section{Previous Work}~\label{prev}

For the case of Linear FSRs (LFSRs), 
there are two main approaches to constructing an LFSR with the degree of parallelization $p$: (1) synthesis of subsequences representing $p$ decimation of some phase shift of the original LFSR sequence and (2) computation of the set of states reachable from any state in $p$ steps.

Let  $S$ be a sequence produced by an LFSR whose characteristic polynomial $g(x)$ of degree $n$ is irreducible in $GF(2)$. Let $\alpha$ be a root of $g(x)$ and let $T$ be the period of $S$. 
In the method based of synthesis of subsequences~\cite{Lempel1971}, the sequence $S$ is decomposed into $p$ subsequences $S_p^j$, each
representing a $p$ decimation of $j$th phase shift of $S$. In other words, the $i$th element of  $S_p^j$ is equal to $i \cdot p + j$ element of $S$. 
By Zierler's theorem~\cite{Zi59}, for $0 \leq j < p$, the subsequences $S_p^j$ can be generated by an LFSR with the following properties:
\begin{itemize}
\item The minimum polynomial of $\alpha^d$ in $GF(2^n)$ is the characteristic polynomial $q^*(x)$ of the new LFSR which has:
\begin{itemize}
\item Period $T^* = T/gcd(d,T)$,
\item Degree $n^*$, which is the multiplicative order of 2 in $Z(T^*)$.
\end{itemize}
\end{itemize}
The Berlekamp-Massey algorithm~\cite{Ma69} or its generalizations~\cite{FeT91} can be used to find the smallest LFSR for each subsequence $S_p^j$. 
The size of each LFSR is $n^*$, which is at most $n$, i.e. the overall number of bits in $p$ LFSRs is at most $p \times n$.
This method is applicable to any degree of parallelization $p$ which is not a multiple of the period $T$. 




The second approach is based on computing the set of states reachable from any state in $p$ steps. This is usually done by computing 
$p$th power of the connection matrix of the LFSR~\cite{GoW98,MuS06}. Such an approach is applicable to the degrees of parallelization $1 < p \leq n$. The size of the register with the degree of parallelization $p$ in this case is the same as the size of the original LFSR, $n$.  

For the case of Non-Linear FSRs (NLFSRs), algorithms for finding a shortest NLFSR generating a
given binary sequence have been presented in~\cite{Ja91,RiK05,RiKK05},
and~\cite{LiKK07}. An NLFSR with the degree of parallelization $p$ can be constructed by computing the set of states reachable from any state in $p$ steps, as in the approach (2) for LFSR. This can be done by computing $p$th power of the transition relation of the NLFSR. However, the size of $p$th power of the transition relation of an NLFSR usually grows much faster than in the LFSR case. Therefore, in practice, in applications which use NLFSRs with the degree of parallelization $p$, NLFSRs are selected so that variables of the $p$ left-most stages of the NLFSR are not used in the updating functions. In such a case, an NLFSR with the degree of parallelization $p$ can be constructed by duplicating the updating functions $p$ times~\cite{hell-grain,canniere-trivium,cryptoeprint:2005:415}. 

For binary machines with the degree of parallelization one, an algorithm for constructing a shortest binary machine generating a
given binary sequence has been presented in~\cite{Du10a}.

\begin{algorithm}[t!]
\caption{Construct an $m$-ary machine
which generates an $m$-ary sequence $A = (a_0, a_1, \ldots, a_k)$ with the degree of parallelization one.}
\label{alg2}
\begin{algorithmic}[1]
\FOR{every $i$  from 0 to $m-1$}
\STATE $N_i := 0$; /*counts the number of digits with value $i \in M$*/
\ENDFOR
\FOR{every $j$  from 0 to $k-1$}
\STATE $N_{a_j} := N_{a_j} + 1$; 
\ENDFOR
\STATE $N_{max} := max_{i \in M} N_i$
\FOR{every $i$  from 0 to $m-1$}
\STATE ${\bf B}_i := \emptyset$
\FOR{every $j$  from 0 to $N_{max}-1$}
\STATE ${\bf B}_i := {\bf B}_i \cup \{ j*m + i \}$;  
\ENDFOR
\ENDFOR
\FOR{every $i$  from 0 to $m-1$}
\STATE $B_i := [b_{i,0}, b_{i,1}, \ldots, b_{i,N_{max}-1}]$ is an arbitrary permutation of ${\bf B}_i$;
\STATE $r_i := 0$; /*records how many elements of $B_i$ were used*/
\ENDFOR

\FOR{every $j$  from 0 to $k-1$}
\STATE $s_j := b_{a_j,r_{a_j}}$;  /*$b_{a_j,r_{a_j}}$ is the $r_{a_j}$th element of $B_{a_j}$*/
\STATE $r_{a_j} := r_{a_j}  + 1$;
\ENDFOR 
\STATE $n =  \lceil log_m N_{max} \rceil + 1$; 
\FOR{every $j$  from 0 to $k-1$}
\STATE Expand $s_j$ as an $m$-ary vector $s_j := (s_{j_{n-1}}, s_{j_{n-2}}, \ldots, s_{j_0})  \in M^n$;
\ENDFOR \\
/*The resulting sequence $S = (s_0, s_1, \ldots, s_{k-1})$ is interpreted as a sequence of states of an $m$-ary $n$-stage machine*/
\FOR{every $p$  from 0 to $n-1$}
\FOR{every $i$  from 0 to $m-1$}
\STATE $i{\mbox{\em-set}}(f_p)  = \emptyset$;
\ENDFOR
\ENDFOR
\FOR{every $j$  from 0 to $k-1$}
\FOR{every $p$  from 0 to $n-1$}
\STATE $i = s_{(j+1)_p}$;
\STATE $i{\mbox{\em-set}}(f_p) = i{\mbox{\em-set}}(f_p) \cup \{(s_{j_{n-1}},s_{j_{n-2}},\ldots, s_{j_0})\}$;
\ENDFOR
\ENDFOR
\STATE Return $(f_0, f_1, \ldots, f_{n-1})$;
\end{algorithmic}
\end{algorithm}

\section{Synthesis Algorithm} \label{sa1}

The algorithm presented in this section exploits the property of $m$-ary
$n$-stage machines that {\em  any} $m$-ary $n$-tuple can be the next state of 
a given current state. Note that, in the traditional $n$-stage NLFSRs in the Fibonacci configuration~\cite{Golomb_book}, the next state
overlaps with a current state in $n-1$ positions.
NLFSRs in the Galois configuration are more flexible. However, since they do not allow feedforward connections,
their set of possible next states is still restricted to a certain subset of all possible states~\cite{Du09j}.

The input of the algorithm is an $m$-ary sequence $A$ of length $k$.
First, we show how to construct a sequence of integers $S = (s_0, s_1, \ldots, s_{k-1})$
such that $s_j$ mod $m = a_j$ for all $j \in \{0,1,\ldots,k-1\}$.
We count the number of occurrences of
each of digits with the value $i \in M$ in $A$, $N_i$, and determine the largest number of
occurrences, $N_{max} = max_{i \in M} N_i$. 

Let ${\bf B}_i$ be a set consisting of $N_{max}$ non-negative integers of type $j \cdot m + i$
for all $j \in \{0,1,\ldots,N_ {max}-1\}$ and all $i \in M$. 
Let $B_i =  [b_{i,0}, b_{i,1}, \ldots, b_{i,N_{max}-1}]$ be an arbitrary permutation of ${\bf B}_i$.

Initially, for all $i \in M$, we set to zero a counter $r_i$ which counts how many digits of $B_i$ have been used. 
Then, for every $j$ from 0 to $k-1$, we take the $j$th element of the sequence $A$, $a_j$, 
and assign $s_j$ to $r_{a_j}$th element of $B_{a_j}$. It is easy to see from our construction that
$s_j$ mod $m$ is equal to $a_i$.

Let $S = (s_0, s_1, \ldots, s_{k-1})$ be a sequence constructed as described above.
Each integer $s_i \in S$ can be represented as an $m$-ary expansion
$(s_{i_{n-1}}, s_{i_{n-2}}, \ldots, s_{i_0}) \in M^n$ where 
$n$ is the number of $m$-ary digits needed to represent the largest integer of $S$
and $s_{i_0}$ is the least significant  digit of the expansion.
We interpret each $n$-tuple $(s_{i_{n-1}}, s_{i_{n-2}}, \ldots, s_{i_0})$ as a state
of an $m$-ary $n$-stage machine.
By construction, $s_{i_0} = a_i$ for all $i \in \{0,1,\ldots,k-1\}$.

Next, we define a mapping $s_i \mapsto s_{i+1}$, for all $i \in
\{0,1,\ldots,k-1\}$, where $''+''$ is mod $k$. This mapping assigns 
$s_{i+1}$ to be the next state of a current state $s_i$ of an $m$-ary $n$-stage machine. Each of $m^n - k$ remaining
states of the $m$-ary $n$-stage machine are left unspecified. This 
gives us a freedom to specify the updating functions in a way which 
minimizes their circuit complexity.

The $i$-sets of the updating functions 
implementing the resulting mapping are derived as follows.
Initially $i{\mbox{\em-set}}(f_j) = \emptyset$, for all $j \in \{0,1,\ldots,n-1\}$ and all $i \in M$. For every $j$ from 0 to $k-1$, and every $p$ from 0 to $n-1$, if $s_{(j+1)_p} \not= 0$, where $''+''$ is mod $k$, then we add $(s_{j_{n-1}},s_{j_{n-2}},\ldots, s_{j_0})$ to the $i$-set of $f_p$ where $i = s_{(j+1)_p}$.

The algorithm described above is summarized as Algorithm~\ref{alg2}.
Its worst-case time complexity is $O(n \cdot k)$ (assuming $k > m$ which is normally the case).

\begin{theorem} \label{th1}
The Algorithm~\ref{alg2} constructs an $m$-ary $n$-stage machine 
generating an $m$-ary sequence $A$ of length $k$ with the degree of parallelization one where $n$ is given by
\begin{equation} \label{bound}
n =  \lceil log_m N_{max} \rceil + 1,
\end{equation}
where $N_{max} = max_{i \in M} N_i$.
\end{theorem}
{\bf Proof:}  
At the step 7 of the Algorithm~\ref{alg2}, for each $i \in M$, $N_i$ equals to the number of digits with the value $i$ in the sequence $A$. 
From the step 6 of the Algorithm~\ref{alg2} we can conclude that, for each $i \in M$, 
the largest integer $s_ i \in S$ such that $s_i$ mod $m = i$ is equal to $m (N_i - 1) + i$. 
 We need $\lceil log_m N_i \rceil + 1$
$m$-ary digits to express this integer for any $N_i > 0$.  Since $k > 1$, the number of stages in
the $m$-ary $n$-stage machine is
given by  $\lceil log_m N_{max} \rceil + 1$ where $N_{max} = max_{i \in M} N_i$.
\begin{flushright}
$\Box$
\end{flushright}

The Lemma below shows under which conditions that the bound given by~(\ref{bound}) is an exact lower bound.

\begin{lemma} \label{th2}
Given a purely periodic $m$-ary sequence $A_m$ with the period $k$, any 
$m$-ary machine which generates $A_m$ the degree of parallelization one has at least $n$ stages, where
$n$ is given by~(\ref{bound}).
\end{lemma}
{\bf Proof:} 
The existence of an $m$-ary machine with $n = \lceil log_m N_{max} \rceil + 1$ stages which can generate $A_m$
follows from the Theorem~\ref{th1}.  
It remains to prove that no $m$-ary $n'$-stage machine with $n' < n$ can generate $A_m$.

Assume that such a machine exists. 
Then, if $A_m$ is purely periodic and has the period $k$, to be able to generate 
one digit of $A_m$ per clock cycle with the period $k$, the $m$-ary $n'$-stage machine 
must have at least $N_i$ distinct states whose 0the stage has the value $i$.
We need at least  $\lceil log_m N_i \rceil + 1$ $m$-ary stages
to implement the largest of these states for any $N_i > 0$.
So, we can conclude that $n' \geq \lceil log_m N_{max} \rceil + 1$ which 
contradicts the assumption that $n' < n$.
\begin{flushright}
$\Box$
\end{flushright}

As an example, consider the 4-ary sequence from the Introduction section:
\[
A_{4} = (0,3,1,3,0,2,3,2,3,0).
\]
We have $N_{max} = 4$. So:
\[
\begin{array}{l}
{\bf B}_0 = \{0,4,8,12\}, \\
{\bf B}_1 = \{1,5,9,13\}, \\
{\bf B}_2 = \{2,6,10,14\}, \\
{\bf B}_3 = \{3,7,11,15\}.
\end{array}
\]
Suppose we use following permutations of ${\bf B}_i$s: 
\[
\begin{array}{l}
B_0 = [0,4,8,12],\\
B_1 = [1,5,9,13],\\
B_2 = [2,6,10,14],\\
B_3 = [3,7,11,15].\\
\end{array}
\]
Then we get:
\[
S_4 = (0,3,1,7,4,2,11,6,15,8).
\]

Since $N_{max} = 4$, from the Theorem~\ref{th1} we can conclude that the quaternary machine which generates $A$ has 2 stages.
By applying the mapping described in the Algorithm~\ref{alg2} to $S$, we get the following  $i$-sets for the updating functions $f_0$ and $f_1$:
\[
\begin{array}{l}
0{\mbox{\em-set}}(f_1) =  \{(00),(03),(10),(20)\} \\[1mm]
1{\mbox{\em-set}}(f_1) =  \{(01),(13),(23)\} \\[1mm]
2{\mbox{\em-set}}(f_1) =   \{(02),(33)\} \\[1mm]
3{\mbox{\em-set}}(f_1) =   \{(22)\} \\[1mm]
0{\mbox{\em-set}}(f_0) =  \{(13),(20),(33)\} \\[1mm]
1{\mbox{\em-set}}(f_0) =  \{(03)\} \\[1mm]
2{\mbox{\em-set}}(f_0) =  \{(10),(23)\} \\[1mm]
3{\mbox{\em-set}}(f_0) =  \{(00),(01),(02),(12)\}. \\[1mm]
\end{array}
\]
The defining tables of these functions are shown is Figure~\ref{f_k}. The symbol "-" stands for a don't care value.

\begin{figure}[t!]\footnotesize
\begin{center}
\begin{tabular}{|c|cccc|}
\hline
$x_0 \backslash x_1$ & 0 & 1 & 2 & 3 \\ \hline
0 & 0 & 0 & 0 & - \\ \hline
1 & 1 & - & - & - \\ \hline
2 & 2 & 3 & - & - \\ \hline
3 & 0 & 1 & 1 & 2 \\ \hline
\end{tabular}
~~~~~~
\begin{tabular}{|c|cccc|}
\hline
$x_0 \backslash x_1$ & 0 & 1 & 2 & 3 \\ \hline
0 & 3 & 2 & 0 & - \\ \hline
1 & 3 & - & - & - \\ \hline
2 & 3 & 3 & - & - \\ \hline
3 & 1 & 0 & 2 & 0 \\ \hline
\end{tabular}
\vspace*{1mm}
~~~~~Function $f_1(x_0,x_1)$  ~~~~~~~~~~~~~~ Function $f_0(x_0,x_1)$
\caption{Defining table for the updating functions of the 4-ary 2-stage machine in Figure~\ref{bin_mv}(a). The symbol "-" stands for a don't care (unspecified) value.}\label{f_k}
\end{center}
\end{figure}

\begin{figure}[t!]\footnotesize
\begin{center}

\begin{tabular}{|@{}c@{}|@{}c@{}c@{}c@{}c@{}|}
\hline
~$x_{01}x_{00} \backslash x_{11}x_{10}$~ & ~00~ & 01~ & 10~ & 11~ \\ \hline
00 & 0 & 0 & 0 & 0 \\ \hline
01 & 0 & 0 & 0 & 0 \\ \hline
10 & 1 & 1 & 0 & 0 \\ \hline
11 & 0 & 0 & 0 & 1 \\ \hline
\end{tabular}
~
\begin{tabular}{|@{}c@{}|@{}c@{}c@{}c@{}c@{}|}
\hline
~$x_{01}x_{00} \backslash x_{11}x_{10}$~ & ~00~ & 01~ & 10~ & 11~ \\ \hline
00 & 1 & 1 & 0 & 0 \\ \hline
01 & 1 & 0 & 0 & 0 \\ \hline
10 & 1 & 1 & 0 & 0 \\ \hline
11 & 0 & 0 & 1 & 0 \\ \hline
\end{tabular}

\vspace*{1mm}
Function $f_{11}(x_{00},x_{01},x_{10},x_{11})$ ~~  Function $f_{01}(x_{00},x_{01},x_{10},x_{11})$\\[3mm]

\begin{tabular}{|@{}c@{}|@{}c@{}c@{}c@{}c@{}|}
\hline
~$x_{01}x_{00} \backslash x_{11}x_{10}$~ & ~00~ & 01~ & 10~ & 11~ \\ \hline
00 & 0 & 0 & 0 & 0 \\ \hline
01 & 1 & 0 & 0 & 0 \\ \hline
10 & 0 & 1 & 0 & 0 \\ \hline
11 & 0 & 1 & 1 & 0 \\ \hline
\end{tabular}
~
\begin{tabular}{|@{}c@{}|@{}c@{}c@{}c@{}c@{}|}
\hline
~$x_{01}x_{00} \backslash x_{11}x_{10}$~ & ~00~ & 01~ & 10~ & 11~ \\ \hline
00 & 1 & 0 & 0 & 0 \\ \hline
01 & 1 & 0 & 0 & 0 \\ \hline
10 & 1 & 1 & 0 & 0 \\ \hline
11 & 1 & 0 & 0 & 0 \\ \hline
\end{tabular}

\vspace*{1mm}
Function $f_{10}(x_{00},x_{01},x_{10},x_{11})$ ~~ Function $f_{00}(x_{00},x_{01},x_{10},x_{11})$ 

\caption{Defining tables for the updating functions of the binary  4-stage machine in Figure~\ref{bin_mv}(b) for the case when all don't cares are specified to 0.
The pairs $(f_{11},f_{10})$ and $(f_{01},f_{00})$ encode the 4-valued functions $f_1$ and $f_0$ in Figure~\ref{f_k}, respectively.}\label{f_kb}
\end{center}
\end{figure}

Note that, in Lemma~\ref{th2}, we require that $A$ is purely periodic with the period $k$.
The need for the
latter condition is obvious: if $A$ repeats two or more times within the input sequence length $k$ 
given to the Algorithm~\ref{alg2},
then we need less than eq. (\ref{bound}) stages to generate $A$. The  
former condition is necessary because, in the sequence is eventually periodic, 
we might be able to generate is with a binary machine 
with less than eq. (\ref{bound}) stages. 
As an illustration, consider an eventually periodic binary sequence $(1,1,0,0,1,0,1,0,1)$
with pre-period 3 and period 2. 
By using Algorithm~\ref{alg2}, we can construct a binary machine with 4 stages
which repeats this sequence with the period 9.
However, we can also construct a binary machine with 3 stages 
whose state transition graph has a cycle of length 2, corresponding
to the period (0,1) and has a branch implementing (1,1,0) which leads to the cycle.
In some cases, the binary machine constructed by the latter approach might be smaller than the one 
constructed using the Algorithm~\ref{alg2}.

\section{Generation of Binary Sequences} \label{bin}

We can use $m$-ary $n$-stage machines for generating binary sequences by encoding their $m$-ary stages 
and $m$-valued functions using at most $(\lceil log_2 m \rceil \cdot n)$ binary stages and Boolean functions.

\begin{figure}[t!]
\begin{center}
\resizebox{1\columnwidth}{!} {\input{binary_vs_mvl.xfig.pstex_t}}
\caption{(a) A quaternary 2-stage machine with the degree of parallelization one. (b)  The machine from (a) encoded as a binary 4-stage machine with the degree of parallelization two.}\label{bin_mv}
\end{center}
\end{figure}
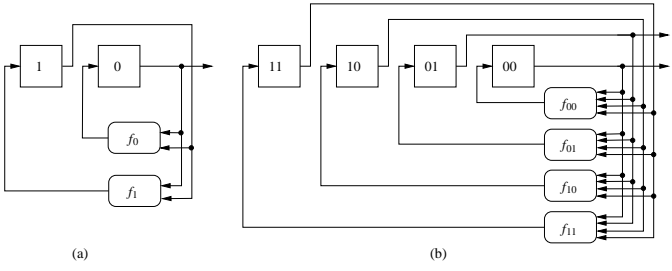

An an example, consider the quaternary 2-stage machine from the example in the previous section.
Figure~\ref{bin_mv}(a) shows its quaternary implementation.
Figure~\ref{bin_mv}(b) shows the same machine in which the updating functions
$f_0$ and $f_1$ are encoded by a pair of Boolean functions $(f_{i0},f_{i1})$, $i \in \{0,1\}$,
using the encoding 0 = (00), 1 = (01), 2 = (10), 3 = (11).  The defining tables for the Boolean functions are 
shown in Figure~\ref{f_kb}. We specified all don't cares of $f_0$ and $f_1$ to 0.
The resulting binary 4-stage machine generates the following sequence $A_{2}$
two bits per clock cycle:
\begin{equation} \label{a2}
A_{2} = (0,0,1,1,0,1,1,1,0,0,1,0,1,1,1,0,1,1,0,0).
\end{equation}

As we showed in the Introduction, if instead of using quaternary encoding, we use Algorithm~\ref{alg2} to construct a
binary machine for $A_{2}$ directly , we get $N_0 = 9$ and $N_1 = 11$ and thus a machine with $n =  \lceil log_2 11 \rceil + 1 = 5$ stages.

Let us see whether we can reduce the number of stages even more is we use 8-are encoding.
We group the bits of $A_{2}$ in triples to get the following 8-ary sequence:
\[
A_8 = (1,5,6,2,7,3,0).
\]
Note that we have added an extra 0 to $A_2$ to make its length a multiple of 3. Using the Algorithm~\ref{alg2} we 
can derive the following sequence of integers $S_8 = (s_0, s_1, \ldots, s_7)$ such that $s_j$ mod $8 = a_j$ for all $j \in \{0,1,\ldots,7\}$:
\[
S_8 = (1,5,6,2,7,3,0).
\]
As we can see, $S = A_8$, because none of the digits of $A_8$ repeat more than once. By the Theorem~\ref{th1},
we need $n =  \lceil log_8 1 \rceil + 1 = 1$ stage to implement this sequence by an 8-ary machine. The updating function of this machine is defined  is Figure~\ref{f_8a}. By encoding the 8-ary 1-stage machine in binary,
 we get a binary 3-stage machine with the updating functions defined  in Figure~\ref{f_8b}
which generates three bits of $A_2$ per clock cycle. So, we gained one more
stage by using the 8-ary encoding. 

Before presenting the main result of the paper, let us formally define $m$-ary encodings.

\begin{definition} \label{enc}
For $m = 2^p$, $p > 0$, an $m$-{\em ary encoding} of a binary sequence $A_2$ of length $k$ is 
the $m$-ary sequence $A_m$ of length $\lceil k/p \rceil$ which is
obtained from $A_2$ by 
replacing the consecutive $p$-tuples of bits of $A_2$, $(a_{i},a_{i+1}, \ldots, a_{i+p-1})$, $i \in \{0,p,2p,\ldots,\lceil k/p \rceil\}$, 
by the value $a_i \cdot m^{p-1} + a_{i+1} \cdot m^{p-2} + \ldots + a_{i+p-1} \cdot m^0$. 
If $k'$ mod $p \not= 0$, then the length of $A$ is extended to the minimum $k'$ such that $k'$ mod $p = 0$ and $k' > k$.
The appended bits are chosen so that the resulting $N_{max} = max_{i \in M} N_i$ is minimum.
\end{definition}

The following theorems gives the lower bound on the number of stages in binary machine with the
degree of parallelization $p$.

 \begin{theorem} \label{t3}
Let $A_2$ be a purely periodic binary sequence with the period $k$. Any binary machine which generates $A_2$ with the degree of parallelization $p \geq 1$ has at least $n$ stages, where $n$ is given by:
\[
n = \lceil log_2 N_{max} \rceil + p
\]
where $N_{max} = max_{i \in M} N_i$ and $N_i$ is to the number of digits with the value $i$ in the 
$m$-ary encoding of $A_2$, $m = 2^p$. 
\end{theorem}
{\bf Proof:}  Let $m = 2^p$ where $p$ is the degree of parallelization, $p > 0$.
From the step 6 of the Algorithm~\ref{alg2} we can conclude that, for each $i \in M$, 
the largest integer $s_ i \in S$ such that $s_i$ mod $m = i$ is equal to $m (N_i - 1) + i$. 
We need $\lceil log_2 N_i \rceil + p$ binary digits to express this integer for any $N_i > 0$ 
Therefore, for $k > 1$, the number of stages in the binary $n$-stage machine is
at most $n \leq \lceil log_2 N_{max} \rceil + p$ where $N_{max} = max_{i \in M} N_i$.

To be able to generate $p$ bits of $A_2$ per clock cycle, the binary $n$-stage machine 
must have at least $N_i$ distinct states whose $p$ lest significant bits correspond to the binary encoding of  
the value $i$. If $A_2$ is purely periodic with the period $k$, we need at least  $\lceil log_2 N_i \rceil + p$ binary stages
to implement the largest of these states for any $N_i > 0$.
Therefore, $n \geq \lceil log_2 N_i \rceil + p$.

So, we can conclude that $n = \lceil log_2 N_{max} \rceil + p$.
\begin{flushright}
$\Box$
\end{flushright}

The technique presented above opens a new possibility for
increasing the throughout  of FSR-based binary sequence generators.
As we mentioned in Section~\ref{prev}, at present,  the generation of $p$-bits of a sequence per clock cycle 
is usually achieved by duplicating the combinatorial logic implementing updating functions of the FSR $p$ times~\cite{hell-grain,canniere-trivium,cryptoeprint:2005:415}. 

As an example, consider the sequence $A_{2}$ given by (\ref{a2}). 
According to the Example V.1 in~\cite{LiKK07}\footnote{The sequence in the Example V.1 in~\cite{LiKK07} does not contain the last bit of $A_{2}$, but this does not change the updating functions of the NLFSR.}, the shortest 
non-linear FSR in the Fibonacci configuration which can generate $A_2$ has 7
stages and the following updating function of the stage 6:
\[
\begin{array}{ll}
f_6 = & \overline{x}_0 \overline{x}_1 \oplus  x_0 \overline{x}_1 \oplus \overline{x}_0 x_1 \oplus x_0 x_1 \overline{x}_2 x_3 \oplus \overline{x}_0 x_1 x_2 x_3 \\[1mm]
 & \oplus x_0 \overline{x}_1 \overline{x}_2 x_3 \oplus \overline{x}_0 x_1 x_2 x_3  \overline{x}_4 x_5  \overline{x}_6 \oplus x_0 x_1 \overline{x}_2 x_3 x_4  x_5 \overline{x}_6. \\
\end{array}
\]
The updating functions of the remaining stages of the NLFSR are of type $f_i = x_{i+1}$, for $i \in \{0,1,\ldots,5\}$.
If we use the number of 2-input XORs and ANDs  
as a measure of cost, then the cost of $f_6$ is 24 ANDs + 7 XORs.

\begin{figure}[t!]\footnotesize
\begin{center}
\begin{tabular}{|c|cccccccc|} \hline
$x_0$ & 0 & 1 & 2 & 3 & 4 & 5 & 6 & 7\\ \hline
	   & 1 & 5 & 7 & 0 & - & 6 & 2 & 3\\ \hline
\end{tabular}
\caption{Defining table for the updating function $f_0$ of the 8-ary 1-stage machine from the example.}\label{f_8a}
\end{center}
\end{figure}

\begin{figure}[t!]\footnotesize
\begin{center}
\begin{tabular}{|c|cccccccc|} \hline
$x_{02}x_{01}x_{00}$ & 000 & 001 & 010 & 011 & 100 & 101 & 110 & 111\\ \hline
	   		& 0 & 1 & 1 & 0 & 0 & 1 & 0 & 0\\ \hline
\end{tabular}  

\vspace*{1mm}
Function $f_{02}(x_{00},x_{01},x_{02})$ \\[3mm]

\begin{tabular}{|c|cccccccc|} \hline
$x_{02}x_{01}x_{00}$ & 000 & 001 & 010 & 011 & 100 & 101 & 110 & 111\\ \hline
	   		& 0 & 0 & 1 & 0 & 0 & 1 & 1 & 1\\ \hline
\end{tabular}

\vspace*{1mm}
Function $f_{01}(x_{00},x_{01},x_{02})$ \\[3mm]

\begin{tabular}{|c|cccccccc|} \hline
$x_{02}x_{01}x_{00}$ & 000 & 001 & 010 & 011 & 100 & 101 & 110 & 111\\ \hline
	   		& 1 & 1 & 1 & 0 & 0 & 0 & 0 & 1\\ \hline
\end{tabular}

\vspace*{1mm}
Function $f_{00}(x_{00},x_{01},x_{02})$ \\[3mm]

\caption{Defining tables for the updating functions $(f_{02},f_{01},f_{00})$ representing the binary encoding of the 8-valued function in Figure~\ref{f_8a} for the case when the don't care is specified to 0.}\label{f_8b}
\end{center}
\end{figure}

On the other hand, as shown above, we can generate 3-bits of $A_2$ per clock cycle using 
the 3-stage binary machine with the
updating functions defined in Figure~\ref{f_8b}. We can
express these functions as follows:
\[
\begin{array}{l}
f_{02} = x_{00} \overline{x}_{01} \oplus \overline{x}_{00} x_{01} \overline{x}_{02} \\
f_{01} = \overline{x}_{00} x_{01} \oplus x_{00} x_{02} \\
f_{00} = \overline{x}_{02} \oplus x_{00} x_{01}. \\
\end{array}
\]
In total,  $f_{02}$,  $f_{01}$ and  $f_{00}$ have 6 AND and 3 XORs. So, the cost of generating 3 bits of $A_2$ per clock cycle using 
this binary 3-stage machine is 3 binary stages of a register + 6 ANDs + 3 XORs. 

Too make a crude comparison of the two costs, let us assume that
the costs of the 2-input AND and the 2-input XOR are 1, and the cost of one stage of a register is 2. Then,
the cost of the NLFSR is 45, while the cost of the binary machine is 15. 
So, the binary machine is not only 3 times faster, but also 3 times smaller.

\begin{table*}[t!]\centering\footnotesize
\begin{tabular}{|c||c|c|c||c|c|c||c|c|} \hline 
 & \multicolumn{3}{c||}{Degree of parallelization = 1} & \multicolumn{3}{c||}{Degree of parallelization = stages in BM} & \multicolumn{2}{c|}{Improvement} \\ \cline{2-9}
Sequence & LFSRs & NLFSRs & BM & LFSRs & NLFSRs & BM & \multirow{2}{*}{$\frac{a4}{a6}$} & \multirow{2}{*}{$\frac{a5}{a6}$} \\ \cline{2-7}
length & $a1$ &  $a2$ &  $a3$ &  $a4$ &  $a5$ &  $a6$ & &  \\  \hline
$2^4$   &       47.35   &       32.93   &       72.38   &       86.3    &       98.68   &       20.85   &       4.73    &       4.14    \\
$2^5$   &       104.33  &       53.4    &       153     &       249.48  &       257.85  &       41.08   &       6.28    &       6.07    \\
$2^6$   &       218.85  &       86.15   &       340     &       654.65  &       1007.62 &       79.03   &       12.75   &       8.28    \\
$2^7$   &       449.03  &       136.18  &       724.5   &       1501.53 &       4081.98 &       151.33  &       26.97   &       9.92    \\
$2^8$   &       885.85  &       236.65  &       1600.1  &       3715.9  &       26638.6 &       371.43  &       71.72   &       10      \\
$2^9$   &       1910.28 &       407.3   &       3258.6  &       8707.73 &       -        &       859.18  &     -    &       10.13   \\
$2^{10}$        &       3889.13 &       757.95  &       7306.78 &       22727.68        &       -        &       1759.3  &    -     &       12.92   \\
$2^{11}$        &       8540.27 &       1399.75 &       15057.5 &      -         &        -       &       3588.9  &    -     &        -       \\
$2^{12}$        &       15664.25        &       2567.45 &       30128.6 &       -        &       -        &       7777.03 &     -    &       -        \\
$2^{13}$        &       30208.38        &       4765.86 &       58946.55        &      -         &      -         &       15719.8 &    -     &      -         \\
$2^{14}$        &      -         &       8817.89 &       114325.91       &      -         &      -         &       32981.89        &     -    &   -            \\
$2^{15}$        &       -        &       16084.3 &       219473.62       &      -         &      -         &       63694.7 &   -      &      -         \\
$2^{16}$        &        -       &       -        &       419118.45       &          -     &       -        &       123947.6        &      -         &      -         \\ \hline
\end{tabular}
\caption{Area results for random sequences (average for 20 sequences);
'-' stands for time out to compute the result (15 min).}
\label{ta1}
\end{table*}

\begin{table*}[t!]\centering\footnotesize
\begin{tabular}{|c||c|c|c||c|c|c||c|c|} \hline 
 & \multicolumn{3}{c||}{Degree of parallelization = 1} & \multicolumn{3}{c||}{Degree of parallelization = stages in BM} & \multicolumn{2}{c|}{Improvement} \\ \cline{2-9}
Sequence & LFSRs & NLFSRs & BM & LFSRs & NLFSRs & BM & \multirow{2}{*}{$\frac{a4}{a6}$} & \multirow{2}{*}{$\frac{a5}{a6}$} \\ \cline{2-7}
length & $a1$ &  $a2$ &  $a3$ &  $a4$ &  $a5$ &  $a6$ & &  \\  \hline
$2^4$   &       49      &       34      &       81.5    &       115     &       155     &       20.5    &       5.61    &       7.56    \\
$2^5$   &       105     &       54.5    &       164     &       241     &       411.5   &       48.5    &       4.97    &       8.48    \\
$2^6$   &       279     &       92.5    &       347.5   &       782     &       6347.5  &       91.5    &       8.55    &       69.37   \\
$2^7$   &       493     &        -         &       707     &       1747    &       -          &       165.5   &       10.56   &    -     \\
$2^8$   &       1093    &        -         &       1486.5  &       4556    &     -            &       470     &       9.69    &    -     \\
$2^9$   &       2161    &        -         &       2737    &       12531   &      -         &       909.5   &       13.78   &    -     \\
$2^{10}$        &       4509    &        -         &       6348.5  &       34660   &      -         &       1865.5  &       18.58   &   -      \\
$2^{11}$        &       9097    &         -        &       11269   &       82954   &      -         &       3874    &       21.41   &   -            \\
$2^{12}$        &       19379   &       -          &       23073.5 &       -        &      -         &       8324.5  &   -      &   -            \\
$2^{13}$        &       36951   &       -          &       39905   &     -          &       -        &       13888   &     -    &    -           \\
$2^{14}$        &       74089   &       -          &       80422.5 &      -         &      -         &       22720.5 &   -     &    -           \\
$2^{15}$        &       -       &         -        &       140433  &      -         &        -       &       43094.5 &   -      &     -          \\
$2^{16}$        &       -        &        -       &       292710.5        &         -      &      -         &       82670   &        -       &       -        \\ \hline
\end{tabular}
\caption{Area results for complementary sequences;
'-' stands for time out to compute the result (15 min).}
\label{ta2}
\end{table*}

\begin{table*}[t!]\centering\footnotesize
\begin{tabular}{|c||c|c|c||c|c|c||c|c|} \hline 
 & \multicolumn{3}{c||}{Degree of parallelization = 1} & \multicolumn{3}{c||}{Degree of parallelization = stages in BM} & \multicolumn{2}{c|}{Improvement} \\ \cline{2-9}
Sequence & LFSRs & NLFSRs & BM & LFSRs & NLFSRs & BM & \multirow{2}{*}{$\frac{a4}{a6}$} & \multirow{2}{*}{$\frac{a5}{a6}$} \\ \cline{2-7}
length & $a1$ &  $a2$ &  $a3$ &  $a4$ &  $a5$ &  $a6$ & &  \\  \hline
17      &       42      &       33      &       68.5    &       84      &       110     &       19      &       4.42    &       5.79    \\
31      &       97      &       44.5    &       146.5   &       281     &       192.5   &       31.5    &       8.92    &       6.11    \\
61      &       231.5   &       83.5    &       311     &       667.5   &       1248.5  &       89.5    &       7.46    &       13.95   \\
127     &       482     &       136.5   &       640.5   &       1901    &       10157.5 &       180.5   &       10.53   &       56.27   \\
257     &       833     &       248     &       1357.5  &       3115    &       20787   &       247     &       12.61   &       84.16   \\
557     &       2144.5  &       408     &       2900    &       9629.5  &       -        &       862.5   &       11.16   &    -     \\
1021    &       3796    &       733     &       6779    &       19369   &      -         &       1906    &       10.16   &    -     \\
2053    &       8016.5  &       1356.5  &       13080   &       47263.5 &      -         &       3652    &       12.94   &     -         \\
4099    &       16358   &       2596.5  &       25491.5 &      -         &       -        &       7654    &    -     &      -         \\
8233    &       33930.5 &        -         &       50691   &       -        &       -        &       16211   &     -    &      -         \\
10223   &       42422   &        -         &       71780   &       -        &       -        &       20160.5 &    -     &     -          \\
16127   &       63012   &        -         &       116037  &      -         &      -         &       32423.5 &   -      &     -            \\ \hline
\end{tabular}
\caption{Area results for extended Legendre sequences;
'-' stands for time out to compute the result (15 min).}
\label{ta3}
\end{table*}

\section{Experimental Results} \label{exp}

To evaluate the presented approach, we compared the areas of binary machines, LFSRs and NLFSRs generating the same sequence for 3 types of sequences: truly random, complementary, and Legendre. 
All experiments were run on a PC with Intel dual-core 1.8 GHz 
processor and 2 Gbytes of memory. The area was computed using ABC synthesis tool~\cite{abc} 
by first optimizing the circuits with {\em resyn} script and then by mapping them with {\em map}.
In the results reported below, 1 unit of area is equal to the area of a 2-input NAND gate.

In the first set of experiments, for each $n$ in the range $4 \leq n \leq 16$, we generated 20 truly random sequences of length $2^n$ using the method~\cite{trully_random}. Columns 2-4 of Table~\ref{ta1} show the areas 
of the resulting LFSRs, NLFSRs and binary machines (BM) 
for the degree of parallelization one.
Columns 5-7 of Table~\ref{ta1} shows similar results for the 
degree of parallelization equal to the number of stages in binary machines (which is always less or equal to the number of stages in LFSRs and NLFSRs).
Each entry is an average for 20 sequences. 

LFSRs are quite bad for generating truly random sequences.\footnote{Note that there is a subset of pseudo-random sequences, called $m$-sequences,
for which LFSRs are extremely efficient. An $n$-stage LFSR
with a primitive polynomial of degree $n$ generates an $m$-sequence of length $2^n-1$.
If the primitive polynomial has $k$ non-zero terms, then to implement such an LFSR 
with the degree of parallelization $p$, we need $n$ stages and no more than $k * p$ XORs. 
However, due to the linearity of LFSRs $m$-sequences they are easy to reconstruct from a short segment.} 
The number of
their stages grows roughly as a half of the sequence length.  
For NLFSRs, the number of stages grows much slower. However, the combinatorial area of
parallel NLFSRs grows so fast that they become 
hard to synthesize for random sequences longer than 256 bits. As we can see from Table~\ref{ta1}, on average, the area of parallel binary machines is an order of magnitude smaller than the area of parallel LFSRs and NLFSRs.
 
Table~\ref{ta2} shows the results for complementary sequences. {\em Complementary}
sequences are a pair of sequences whose aperiodic autocorrelation coefficients sum up to zero~\cite{golay}. These sequences are known to have a tightly low peak-to-mean envelope power ratio, 
good error detection capabilities, and high nonlinearity~\cite{DaJ98}. They are recommended for orthogonal frequency division multiplexing~\cite{DaJ98} and for multicarrier code division multiple access systems~\cite{popovich}
We can see that, on average, parallel binary machines are an order of magnitude smaller than parallel LFSRs and NLFSRs.

Table~\ref{ta3} shows the results for extended Legendre sequences. Extended Legendre 
sequences  are known to have the asymptotic merit factor of 6.3421, which is the highest of all known families of sequences of an arbitrary length~\cite{KrP04}. The higher the merit factor of a sequence which is used to modulate a signal, the more uniformly the signal energy is distributed over the frequency range. This is important for spread-spectrum communication systems, ranging systems, and radar systems~\cite{popovich,KrP04}. Again, on average, parallel binary machines are an order of magnitude smaller than parallel LFSRs and NLFSRs.

\section{Conclusion} \label{con}

In this paper, we 
present a method for constructing binary machines with the minimum number of stages for a given degree of parallelization.
Our experimental results show that, for sequences with high linear complexity, such as complementary, 
Legendre, or truly random sequences, parallel binary machines are an order of magnitude smaller 
than parallel LFSRs and NLFSRs generating the same sequence. 

Our results can be beneficial for any application which requires sequences with high spectrum efficiency or high security, such as data transmission, wireless communications, and cryptography.

\section{Acknowledgments}
This work was supported in part by a research grant 621-2010-4388 from the Swedish 
Research Council.


\end{document}

%% file: binary_machine.xfig.pstex_t
\begin{picture}(0,0)%
\includegraphics{binary_machine.xfig.pstex}%
\end{picture}%
\setlength{\unitlength}{3947sp}%
\begingroup\makeatletter\ifx\SetFigFont\undefined%
\gdef\SetFigFont#1#2#3#4#5{%
  \reset@font\fontsize{#1}{#2pt}%
  \fontfamily{#3}\fontseries{#4}\fontshape{#5}%
  \selectfont}%
\fi\endgroup%
\begin{picture}(6354,3672)(1789,-5923)
\put(5193,-3577){\makebox(0,0)[lb]{\smash{{\SetFigFont{12}{14.4}{\rmdefault}{\mddefault}{\updefault}{\color[rgb]{0,0,0}$0$}%
}}}}
\put(6451,-4120){\makebox(0,0)[lb]{\smash{{\SetFigFont{12}{14.4}{\rmdefault}{\mddefault}{\updefault}{\color[rgb]{0,0,0}$f_0$}%
}}}}
\put(6376,-4786){\makebox(0,0)[lb]{\smash{{\SetFigFont{12}{14.4}{\rmdefault}{\mddefault}{\updefault}{\color[rgb]{0,0,0}$f_{n-2}$}%
}}}}
\put(6376,-5386){\makebox(0,0)[lb]{\smash{{\SetFigFont{12}{14.4}{\rmdefault}{\mddefault}{\updefault}{\color[rgb]{0,0,0}$f_{n-1}$}%
}}}}
\put(3451,-3586){\makebox(0,0)[lb]{\smash{{\SetFigFont{12}{14.4}{\rmdefault}{\mddefault}{\updefault}{\color[rgb]{0,0,0}$n-2$}%
}}}}
\put(2251,-3586){\makebox(0,0)[lb]{\smash{{\SetFigFont{12}{14.4}{\rmdefault}{\mddefault}{\updefault}{\color[rgb]{0,0,0}$n-1$}%
}}}}
\put(3151,-4036){\makebox(0,0)[lb]{\smash{{\SetFigFont{12}{14.4}{\rmdefault}{\mddefault}{\updefault}{\color[rgb]{0,0,0}$n$ register stages}%
}}}}
\end{picture}%

%% file: binary_vs_mvl.xfig.pstex_t
\begin{picture}(0,0)%
\includegraphics{binary_vs_mvl.xfig.pstex}%
\end{picture}%
\setlength{\unitlength}{3947sp}%
\begingroup\makeatletter\ifx\SetFigFont\undefined%
\gdef\SetFigFont#1#2#3#4#5{%
  \reset@font\fontsize{#1}{#2pt}%
  \fontfamily{#3}\fontseries{#4}\fontshape{#5}%
  \selectfont}%
\fi\endgroup%
\begin{picture}(9609,3745)(3079,-3869)
\put(3541,-1111){\makebox(0,0)[lb]{\smash{{\SetFigFont{12}{14.4}{\rmdefault}{\mddefault}{\updefault}{\color[rgb]{0,0,0}$1$}%
}}}}
\put(4666,-1111){\makebox(0,0)[lb]{\smash{{\SetFigFont{12}{14.4}{\rmdefault}{\mddefault}{\updefault}{\color[rgb]{0,0,0}$0$}%
}}}}
\put(4851,-2141){\makebox(0,0)[lb]{\smash{{\SetFigFont{12}{14.4}{\rmdefault}{\mddefault}{\updefault}{\color[rgb]{0,0,0}$f_0$}%
}}}}
\put(4844,-2904){\makebox(0,0)[lb]{\smash{{\SetFigFont{12}{14.4}{\rmdefault}{\mddefault}{\updefault}{\color[rgb]{0,0,0}$f_1$}%
}}}}
\put(6901,-1111){\makebox(0,0)[lb]{\smash{{\SetFigFont{12}{14.4}{\rmdefault}{\mddefault}{\updefault}{\color[rgb]{0,0,0}$11$}%
}}}}
\put(8026,-1111){\makebox(0,0)[lb]{\smash{{\SetFigFont{12}{14.4}{\rmdefault}{\mddefault}{\updefault}{\color[rgb]{0,0,0}$10$}%
}}}}
\put(9151,-1111){\makebox(0,0)[lb]{\smash{{\SetFigFont{12}{14.4}{\rmdefault}{\mddefault}{\updefault}{\color[rgb]{0,0,0}$01$}%
}}}}
\put(10276,-1111){\makebox(0,0)[lb]{\smash{{\SetFigFont{12}{14.4}{\rmdefault}{\mddefault}{\updefault}{\color[rgb]{0,0,0}$00$}%
}}}}
\put(11101,-1636){\makebox(0,0)[lb]{\smash{{\SetFigFont{12}{14.4}{\rmdefault}{\mddefault}{\updefault}{\color[rgb]{0,0,0}$f_{00}$}%
}}}}
\put(11101,-2236){\makebox(0,0)[lb]{\smash{{\SetFigFont{12}{14.4}{\rmdefault}{\mddefault}{\updefault}{\color[rgb]{0,0,0}$f_{01}$}%
}}}}
\put(11101,-2836){\makebox(0,0)[lb]{\smash{{\SetFigFont{12}{14.4}{\rmdefault}{\mddefault}{\updefault}{\color[rgb]{0,0,0}$f_{10}$}%
}}}}
\put(11101,-3436){\makebox(0,0)[lb]{\smash{{\SetFigFont{12}{14.4}{\rmdefault}{\mddefault}{\updefault}{\color[rgb]{0,0,0}$f_{11}$}%
}}}}
\end{picture}%